\documentstyle[12pt]{article}
\newcommand{\bb}{\begin{eqnarray}}
\newcommand{\ee}{\end{eqnarray}}

\newcommand{\m}{\mu}
\newcommand{\n}{\nu}

\newcommand{\pl}{\partial}

\newcommand{\bR}{{\bf R}}
\newcommand{\bZ}{{\bf Z}}

\newcommand{\bt}{\bar{\tau}}
 
\begin{document}
\begin{titlepage}
\vspace{-.3cm}
 \hspace*{\fill} {\small
hep-th/9508159 }\\
\hspace*{\fill} {\small NTUA-95/51} \\
\hspace*{\fill} {\small August 1995}\\
\vspace{.6cm}
\begin{center}
{\LARGE A CANONICAL APPROACH TO  S-DUALITY IN

ABELIAN GAUGE
THEORY}\\
\vspace{1.2cm}

{\large Alexandros A. Kehagias}
\footnote{Address after September 1995: Phys. Dept., Inst. f\"ur Theor.
Phys., Technische Univ., M\"unchen, D-85748 Garching, Germany}\\
\vspace{-.2cm}
{\large Physics Department \\
\vspace{-.2cm}
National Technical University\\
\vspace{-.2cm}
15780 Zografou Campus Athens, Greece\\
E-Mail: kehagias@sci.kun.nl}\\
\vspace{1.3cm}

{\large Abstract}
\end{center}
\vspace{.7cm}

We examine the electric-magnetic duality for a U(1) gauge theory on a
general four manifold which
generates the  $SL(2,{\bf Z})$ group. The partition functions for such
a theory transforms as a modular form of specific weight.
 However, in the canonical approach,
 we show that  S-duality for the abelian theory,
like T-duality, is  generated by  a canonical
transformation leading to a modular invariant partition function.

\end{titlepage}

\newpage

According to the S-duality assertion, certain four-dimensional gauge
theories with a $\theta$-term are invariant under $SL(2,\bZ)$ modular
transformations on the
complex coupling
\bb
\tau=\frac{\theta}{2\pi}+i\frac{4\pi}{g^2}\,  , \label{T}
\ee
where $\theta$, $g$ are the theta angle and the coupling constant,
respectively \cite{1}. This assertion has been tested for N=4 supersymmetric
Yang-Mills theory  by calculating the corresponding partition
function \cite{2} and the result was in favour of S-duality
if the gauge group is replaced
by its dual.
For an abelian $U(1)$ gauge theory, S-duality
extends the electric-magnetic duality \cite{2'},\cite{2''}
of electromagnetism on ${\bR}^4$ to the full action of $SL(2,\bZ)$
for a general four-manifold.
The corresponding partition function fails to be modular invariant but
rather it transforms as a modular form of specific weight
\cite{3}--\cite{5}.
This can  be seen by compactifing $M$ over a torus.
The reduced two-dimensional theory looks like a linear
sigma-model where the internal components of the photon are  scalar
fields. S-duality is then reduced to T-duality \cite{6} and the
modular anomaly appears because there is no dilaton field to
compensate it \cite{6'},\cite{6''}.
However, when the four manifold is of the form ${\bf R}\!\times\!X$
and a time direction can be chosen, by employing the canonical
approach we will see that S-duality, like T-duality \cite{7},\cite{8},
 is just a canonical
transformation leading to a  modular invariant partition function.

Let us  consider a $U(1)$ gauge field $A_\m$, i.e., a connection on a
line bundle L over a four dimensional manifold $M$ with corresponding
 field strength $F_{\m\n}=\pl_\m A_\n-\pl_\n A_\m$.
This theory is described by the  action
\bb
S[A]= \frac{1}{e^2}\int_{M} F\wedge \ast F
-i\frac{\theta}{8\pi^2}\int_{M} F\wedge F\, , \label{ac}
\ee
where $F=\frac{1}{2}F_{\m\n}dx^\m\wedge dx^\n$ and $\ast
F=\frac{1}{4}F^{\m\n}\epsilon_{\m\n\rho\kappa}dx^\rho\wedge
dx^\kappa$. We  assume that
$M$ is Euclidean  so that,
the saddle points of (\ref{ac})
 turns out to be the  Hodge-de Rham equations
\bb
dF=0 &,& d\ast\! F=0\, ,
\ee
for harmonic two-forms.
In general,  the number
 of harmonic n-forms on $M$ is $b_n$ where $b_n$ is the $n^{th}$
Betti number, i.e., the dimension of the $n^{th}$ de Rham cohomology
group $H_2(M,{\bf R})$ while the Euler number of $M$
is given by the algebraic sum
$\chi=\sum_n (-1)^nb_n$.

  The partition function of the $U(1)$ gauge theory defined on the
four-manifold $M$ is given by
\bb
Z=C\frac{1}{|\cal{G}|}\int_{M}
 DA e^{-S[A]}\, , \label{Z}
\ee
where C is a regularization constant, the  integration is
 over all the $U(1)$ gauge fields on $M$ and we
divide by the volume $|\cal{G}|$ of the gauge group as usual.
Moreover, a sum over isomorphic classes of the line bundle $L$ over
$M$ is understood.
By employing the Faddeev-Popov procedure to factor out this volume,
the partition function may easily be evaluated to be
\bb
Z=
C\frac{det^\prime
e^{-1}\Delta_{FP}}{(det^\prime e^{-2}\Delta_A)^{1/2}} Z_{cl}\,,
\ee
where
\bb
Z_{cl}=\sum_{saddle\, \, points} e^{-S[A_{cl}]}\, ,
\ee
is the sum of all saddle-point  configurations of the action
(\ref{ac}).
 $det^\prime e^{-1}\Delta_{FP},\, det^\prime e^{-2}\Delta_A$
 denote the Faddeev-Popov
determinant and the determinant of the kinetic term for the gauge
field (without counting the zero modes).
The former is the determinant of the scalar Laplacian $\Delta_0$ while
the latter of the Laplacian $\Delta_1$ for one-forms.
One may easily verify that
\bb
det^\prime e^{-2}\Delta_A=
(\frac{1}{e^2})^{\zeta_{\Delta_A}(0)}
det^\prime \Delta_A\, , \label{AA}
\ee
where $\zeta_\Delta(s)$ is the generalised zeta function for the
operator $\Delta$. Since  $\zeta_{\Delta}(0)$ is just the dimension
of  $\Delta$ (infinite if not properly regularized)
 without counting the zero modes, we may write
\bb
\zeta_{\Delta_A}(0)=dim\Delta_1-b_1\, ,
\ee
 Similarly, we have
\bb
det^\prime e^{-1}\Delta_{FP}=
(\frac{1}{e^2})^{\frac{1}{2}\zeta_{\Delta_{FP}}(0)}det^\prime\Delta_{FP}
\, , \label{FPD}
\ee
with
\bb
\zeta_{\Delta_{FP}}(0)=dim\Delta_0-b_0\, . \label{FP}
\ee
Thus, by choosing the  constant C in eq.(\ref{Z}) to be
\bb
C=
(Im\tau)^
{\frac{1}{2}(dim\Delta_1-\dim\Delta_0)}(4\pi)^{\frac{1}{2}(dim\Delta_0
-dim\Delta_1+b_1-b_0)}\, , \label{C}
\ee
the partition function is properly regularized and
may be  written as
\bb
Z=
(Im\tau)^{\frac{1}{2}(b_1-b_0)}
\frac{det^\prime\Delta_{FP}}{(det^\prime\Delta_A)^{1/2}} Z_{cl}\, ,
\label{Z1}
\ee
where numerical factors have been omitted.
Note that we may express the action
(\ref{ac}) in terms of self-dual ($F^+=F+\ast F$),
anti-self-dual ($F^-=F-\ast F$) fields and the complex
coupling $\tau$ defined in eq.(\ref{T}) as
\bb
S=\frac{i}{16\pi}\int_{M} (\bt F^-\wedge F^--\tau F^+\wedge F^+).
\label{action}
\ee
 Let us suppose that in $M$ there
exist non-trivial two-cycles, that is
closed surfaces which do not bound any 3-dimensional
sub-manifold of $M$. One may then consider the flux of
 the Maxwell field strength $F_{\m\n}$, through these cycles which
 satisfies the Dirac quantization condition
\bb
\int_{S_I}F=2\pi n^I\, ,\label{Dirac}
\ee
where $n^I\in {\bf Z}$ and $I=1,...,b_2$.
We may define a basis $\alpha_I$
 of harmonic two-forms ($d\alpha_I=d\!\ast\!\alpha_I=0$)
normalised as $\int_{S_I}\alpha_J=\delta_{IJ}$ which generates the
harmonic representatives in $H^2(M,{\bf R})$ and allows us to express
the field strength $F$ as
\bb
F=2\pi\sum_I n^I\alpha_I\, . \label{Di}
\ee
 We may also define a basis in the
space  of self-dual/anti-self-dual harmonic two-forms by
\bb
\alpha_I^{\pm}=\alpha_I\pm\ast\alpha_I\, .
\label{sd}
\ee

By employing eqs.(\ref{Di},\ref{sd}), we may  express the action
(\ref{action}) as
\bb
S[A_{cl}]=
i\frac{\pi}{4}  \bt n^IH_{IJ}^+n^J -i\frac{\pi}{4}
\tau n^IH_{IJ}^-n^J\, ,\label{ac1}
\ee
where
\bb
H_{IJ}^\pm=\int_{M^4} \alpha^\pm\wedge\alpha^\pm\, ,\label{met}
\ee
is the intersection form for harmonic
self-dual and anti-self-dual two-forms.
The partition function in eq.(\ref{Z1}) may then be written as
\bb
Z=
(Im\tau)^{\frac{1}{2}(b_1-b_0)}
\frac{det^\prime\Delta_{FP}}{(det^\prime\Delta_A)^{1/2}}
 \sum_{n^I}
e^{i\frac{\pi}{4} \tau n^IH_{IJ}^-n^J
-i\frac{\pi}{4}  \bt n^IH_{IJ}^+n^J}\,.
\label{ZZ}
\ee

To examine S-duality transformations which are generated by
$T:\,\tau \rightarrow \tau+1$  and
$S:\, \tau\rightarrow -1/\tau$,
let us recall  one of the basic invariants of a 4-manifold $M$, the
intersection form $\omega$.  If $M$ has a smooth structure, it is defined by
using the de Rham cohomology  $H^*(M)$  as
\bb
\omega(\alpha,\beta)=\int_{M} \alpha\wedge\beta \, , \label{om}
\ee
for $\alpha,\, \beta \in H^2(M,{\bf R})$ \cite{M}. The dimensionality of
$\omega$ is $b_2$ and the number of its positive (negative)
eigenvalues is $b_2^+ (b_2^-)$. The signature $\sigma$ of $\omega$ is
defined as the number of positive eigenvalues minus the negative ones,
i.e., $\sigma=b_2^+-b_2^-$ and it is a topological invariant. For a
simply-connected spin manifold, $\omega(\alpha,\alpha)$ is an even
integer, otherwise is odd. Thus, for a spin manifold by taking
$\alpha=F/2\pi$ one may verify that the action (\ref{ac}) is invariant
under $\tau\rightarrow \tau+1$ while for a non-spin manifold
(\ref{ac}) is invariant under $\tau\rightarrow\tau+2$ \cite{2}.

To examine the transformation properties of the partition function
under $\tau\rightarrow-1/\tau$, let us consider the dual theory which
can be found from the first-order action
\bb
\tilde{S}=\frac{1}{g^2}\int G\wedge \ast G-i\frac{\theta}{8\pi^2}\int G\wedge
G +\frac{i}{2\pi} dG\wedge B \, , \label{DA}
\ee
where $G=\frac{1}{2}G_{\m\n}dx^\m\wedge dx^\n$ and $B=B_\m dx^\m$  are
a 2-form and a 1-form, respectively. It can be constructed by
requiring invariance under gauge transformations of the ``third
kind":$A\rightarrow A+B$ \cite{p}. The partition function for this
theory is given by
\bb
Z=C \int_M \frac{DB}{|\tilde{{\cal G}}|}DGe^{-\tilde{S}}, \label{DZ}
\ee
where we have divide by the volume
$|\tilde{{\cal G}}|$ of the gauge group which transforms $B$ according
to $B\rightarrow B+d\Lambda$. By
integrating out the Lagrange multiplier field $B$ on  a topologically
trivial manifold $M$, we get $dG=0$ which implies that $G=d\tilde{A}$
and  the original $U(1)$ gauge theory described by (2) is recovered.
However, in a non-trivial manifold $M$, in order $G$ to be  the
curvature (field strength) of a $U(1)$ connection (gauge field), the
Dirac quantization condition must be taken into account, i.e.,
eq.(\ref{DZ}) must be implemented by the sum
\bb
\sum_{n^I\in {\cal Z}}\delta\left(n^I-\int_{S_I}\frac{G}{2\pi}\right)=
\sum_{m^I\in {\cal Z}} e^{-im^I\int_{S_I}G} \, ,
\ee
where the sum in the righthand side in the above equation is over the
dual lattice.
We may also integrate the G-field and for this it is more
convinient to express (\ref{DA}) as
\bb
\tilde{S}=\frac{i}{16\pi}\int_M \left(\bt G^-\wedge G^--\tau G^+\wedge
G^+\right)-\frac{i}{8\pi}\int_M\left( G^+\wedge W^++G^-\wedge
W^-\right) \, ,
\ee
where $W=dB$ and  $G^\pm=G\pm\ast G$,
$W^\pm=W\pm\ast W$ are the self-dual and
anti-self-dual parts of $G$ and $W$. Since self-dual and anti-self-dual
forms are orthogonal, we may integrate them separately. The result is,
ignoring numerical factors,
\bb
Z(\tau)=C\bt^{-\frac{1}{2}B_2^-}\tau^{\frac{1}{2}B_2^+}\int
\frac{DB}{|\tilde{{\cal G}}|}e^{-\tilde{S}} \, , \label{a}
\ee
where
\bb
\tilde{S}=
\frac{i}{16\pi}\int_M \left( (-\frac{1}{\bt})W^-\wedge W^--
(-\frac{1}{\tau})W^+\wedge W^+\right) \, , \label{xx}
\ee
and $B_2^\pm$ is the number of self-dual, anti-self-dual 2-forms on
$M$ in a lattice regulatization. Note that (\ref{xx}) does describe a
$U(1)$ gauge theory since by integrating out $G$ we also get
\bb
\int_{S_I}\frac{W}{2\pi}=m^I\in {\cal Z} \, , \nonumber
\ee
for the fluxes of the dual theory. Thus, eq.(\ref{a}) is indeed the
partition function for a $U(1)$ gauge theory.
As  follows now from eqs. (4,11),
\bb
Z(-\frac{1}{\tau})=C(\tau\bt)^{-\frac{1}{2}(B_1-B_0)}\int
\frac{DA}{|{\cal G}|}
e^{-
\frac{i}{16\pi}\int_M \left( (-\frac{1}{\bt})F^-\wedge F^--
(-\frac{1}{\tau})F^+\wedge F^+\right)} \, , \label{b}
 \ee
where $B_1=dim\Delta_1, B_0=dim\Delta_0$.
By comparing eqs.(\ref{a},\ref{b}) we get
\bb
Z(-\frac{1}{\tau})&=&\tau^{\frac{1}{2}(B_0-B_1+B_2^+)}\bt^{\frac{1}{2}
(B_0-B_1+B_2^-)}Z(\tau) \\
&=&\tau^{\frac{1}{4}(\chi-\sigma)}\bt^{\frac{1}{4}(\chi+\sigma)}Z(\tau)
\, , \label{ZT}
\ee
and therefore $Z(\tau)$ transforms as a modular form of weight
$(\frac{1}{4}(\chi-\sigma),\frac{1}{4}(\chi+\sigma))$.

Let us now examine  S-duality in the Hamiltonian approach.
We will assume that the four-manifold $M$  is Lorentzian and  by
separating space and time, $M$ turns out to be of the
form ${\bf R}\!\times\!X$ where $X$ is a three-manifold. We will also assume
that  $M$ is endowed with the product metric
\bb
ds^2=-dt^2+g_{ij}dx^idx^j\, ,
\ee
where $g_{ij}$ is the metric on $X$.
In this case the action (\ref{ac}) turns out to be
\bb
S=\int dtd^3x\sqrt{g}\left(
-\frac{1}{2e^2}F_{\m\n}F^{\m\n}+\frac{\theta}{32\pi^2}
\epsilon^{\m\n\rho\lambda}F_{\m\n}F_{\rho\lambda}\right)\, ,
\ee
which is of course real.
The canonical momenta are easily found to be
\bb
\pi^i=\frac{2}{e^2}F_{0j}g^{ij}+\frac{\theta}{8\pi^2}\epsilon^{ijk}F_{ij}
\, ,
\ee
and by performing a Legendre transform,  the Hamiltonian is given
by
\bb
H=\frac{e^2}{4}\pi_i\pi^i +\pi^i\partial_i
A_0-\frac{e^2\theta}{16\pi^2}\pi_i\epsilon^{ijk}F_{jk}
+(\frac{1}{2e^2}+\frac{e^2\theta^2}{128\pi^4})F_{ij}F^{ij}\, ,
\label{Ham}
\ee
while the symplectic structure is provided by the equal-time Poisson
bracket
\bb
\{A_i(x),\pi^j(y)\}=\delta_i^j\delta^{(3)}(x-y)\, .
\ee
The partition function is given by
\bb
Z={N}\int DA D\pi\frac{1}{|{\cal G}|}e^{i\int dtd^3x\sqrt{g}(\pi^i
\dot{A}_i-H)} \, , \label{FPI}
\ee
where $N$  is a regularization constant and we have divide as usual
 by the
volume of the $U(1)$ gauge group. It contains a
sum over the isomorphic classes of the line bundle L over $X$ and in each
class we have to integrate over the momenta $\pi^i$ and $A_\m$.
Every $\pi^i$ integration produces  a factor $(1/e^2)^{1/2}\sim
Im\tau^{1/2}$  while
every $A_i$ one produces a factor $(\frac{1}{2e^2}+
\frac{e^2\theta^2}{128\pi^4})^{-1/2}\sim (Im\tau/\tau\bt)^{1/2}$
($A_0$  integrations do not
produce any factors but simply a delta function).
By using a lattice regularization, the number of $A_i$ integrations
is $B_1-B_0$ where $B_1,B_0$ is the number of 1-forms and 0-forms
on $X$, respectively. Since the number of $\pi^i$ integrations equals
the number of $A_i$ ones, the cut-off dependent term coming from the
integration is
\bb
Im\tau^{\frac{1}{2}(B_1-B_0)} Im\tau^{\frac{1}{2}(B_1-B_0)}
(\tau\bt)^{-\frac{1}{2}(B_1-B_0)} \, .
\ee
Thus, a properly regularized, cut-off independent
 partition function is given by
\bb
Z(\tau)= Im\tau^{B_0-B_1}
(\tau\bt)^{\frac{1}{2}(B_1-B_0)}
\int DA D\pi\frac{1}{|{\cal G}|}e^{i\int dtd^3x\sqrt{g}(\pi^i
\dot{A}_i-H)} \, . \label{FPi}
\ee
As usual, $A_0$ has no kinetic term and it is just a Lagrange
multiplier leading to the  constraint
\bb
\nabla_i\pi^i=0\, ,
\label{constr}
\ee
where $\nabla$ is the covariant derivative on $M$.
Let us now perform a canonical transformation generated by
\bb
G=\frac{1}{4\pi}
\int_X d^3x\sqrt{g}(\tilde{A}_i\epsilon^{ijk}F_{jk}+A_i\epsilon^{ijk}
\tilde{F}_{jk}) \, ,
\ee
so that
\bb
\pi^i&=\frac{\delta
G}{\delta A_i}=&\frac{1}{4\pi}\epsilon^{ijk}\tilde{F}_{jk}, \label{1}
\\
\tilde{\pi}^i&=-\frac{\delta
G}{\delta\tilde{A}_i}=&-\frac{1}{4\pi}\epsilon^{ijk}F_{jk}\, .\label{2}
\ee
It follows from (\ref{2}) that the dual momenta $\tilde{\pi}^i$ satisfy
the constraint
\bb
\nabla_i\tilde{\pi}^i=0\, , \label{cost1}
\ee
which can be incorporated in the  dual Hamiltonian by means of a
Langrange multiplier $\tilde{A}_0$.
Therefore,  the dual Hamiltonian is
\bb
\tilde{H}&=&
16\pi^2(\frac{1}{2e^2}+\frac{e^2\theta^2}{128\pi^2})\tilde{\pi}_i
\tilde{\pi}^i+\frac{e^2\theta}{16\pi^2}
\tilde{\pi}^i\epsilon^{ijk}\tilde{F}_{jk}\nonumber \\
&&+\frac{e^2}{32\pi^2}\tilde{F}_{ij}\tilde{F}^{ij}
+\frac{1}{4\pi}\epsilon^{ijk}\tilde{F}_{jk}\partial_i A_0+\tilde{A}_0
\nabla_i\tilde{\pi}^i\, . \label{dH}
\ee
The integration of $A_0$ gives the constraint
\bb
\epsilon^{ijk}\nabla_i\tilde{F}_{ijk}=0\, , \label{cost2}
\ee
which is just  the Bianchi identity and  allows us to express
$\tilde{F}_{ij}$ locally as
\bb
\tilde{F}_{ij}=\partial_i\tilde{A}_j-\partial_j\tilde{A}_i\, .
\ee
However, such expression will fail in general to hold globally in view of
possible non-zero magnetic fluxes
\bb
\int_{S_I}\epsilon^{ijk}\tilde{F}_{ij}dS_k=4\pi n^I\, , \label{DH}
\ee
through non-trivial two-cycles of $X$ according to the Dirac condition.
By performing an inverse Legendre transform in (\ref{dH}),  the
dual Lagrangian  is found to be
\bb
\tilde{L}=-\frac{1}{2\tilde{e}^2}\tilde{F}_{\m\n}\tilde{F}^{\m\n}
+\frac{\tilde{\theta}}{32\pi^2}\epsilon^{\m\n\rho\lambda}
\tilde{F}_{\m\n}\tilde{F}_{\rho\lambda}\, , \label{dL}
\ee
where
\bb
\tilde{e}^2&=&e^2(\frac{\theta}{4\pi^2}+\frac{16\pi^2}{e^4}) \, ,
\nonumber \\
\tilde{\theta}&=&
-\theta(\frac{\theta}{4\pi^2}+\frac{16\pi^2}{e^4})^{-1} \, .
\label{et}
\ee
Thus, the classical dual theory is related to the original one by the
transformation $\tau\rightarrow -1/\tau$.

Since the partition function is invariant under canonical
transformations  we get
\bb
Z(\tau)= Im\tau^{B_0-B_1}
(\tau\bt)^{\frac{1}{2}(B_1-B_0)}
\int D\tilde{A} D\tilde{\pi}\frac{1}{Vol G}e^{i\int dtd^3x\sqrt{g}
(\tilde{\pi}^i
\tilde{\dot{A}}_i-\tilde{H})} \, . \label{FPI1}
\ee
By employing the transformation $\tau\rightarrow-1/\tau$ in
eq. (\ref{FPi})
we get the righthand side of (\ref{FPI1}) so that
\bb
Z(-1/\tau)=Z(\tau) \, , \label{z}
\ee
and therefore the partition function   for the U(1)-gauge theory in the
canonical approach is modular invariant.

Let us also note that there are
no quantum corrections to the generating functional since it is linear
to $A_i$ and $\tilde{A}_i$ \cite{G}. Moreover, physical states  $\phi_k[A_i]$
 and $\psi_k[\tilde{A}_i]$ in the original and in the dual theory
are related through
\bb
\psi_k[\tilde{A}_i]=\int DA e^{iG(A,\tilde{A})}\phi[A_i]\, .
\ee
These states are invariant under gauge transformations, i.e.
$\phi_k[A_i+\partial_i\epsilon]=\phi_k[A_i]$ so that
\bb
G(A-\partial\epsilon,\tilde{A})-G(A,\tilde{A})=2\pi n\, , \, n\in{\bf
Z} \, .
\ee
This condition is satisfied as long as the Dirac quantization
condition (\ref{DH}) is satisfied.

As we have seen above, the partition function in the canonical
approach is modular invariant. On the other hand, in T-duality the
modular anomaly is compensated by a change of the dilaton field
leading to a modular invariant theory. Similarly here, it seems that
the anomaly is compensated by a change of the generalised momenta
$\pi^i$. In this respect, $\pi^i$ integrations imitates metric
integrations in the string case. Finally,
one should expect that the above discussion would also be carried out
for the non-abelian case as well. However, in this case, one cannot
non-trivially satisfy the Gauss' law constraint $D_i\pi^{\alpha
i}=0$ and the dual theory is identical to the original one.
The non-abelian case will be discussed elsewhere.
\vspace{.5cm}

While the present work was being proof-reading, we became aware of
Ref\cite{LO} where  S-duality is also confronted as a canonical
transformation.

\end{document}